\documentclass[showpacs,amsmath,amssymb,10pt,aps,superscriptaddress]{revtex4}
\usepackage[T1]{fontenc}
\usepackage[latin9]{inputenc}
\usepackage{amsmath}
\usepackage{amssymb}
\usepackage{graphicx}
\def\oper{{\mathchoice{\rm 1\mskip-4mu l}{\rm 1\mskip-4mu l}
{\rm 1\mskip-4.5mu l}{\rm 1\mskip-5mu l}}}
\def\<{\langle}
\def\>{\rangle}
\newtheorem{Theorem}{Theorem}
\newtheorem{Lemma}{Lemma}

\newtheorem{Remark}{Remark}

\newtheorem{Proposition}{Proposition}

\usepackage{color}
\usepackage{soul}
\usepackage{cancel}

\newcommand{\LH}{{{\rm L}}(\mathcal{H})}

\makeatletter

\usepackage{tikz}
\usepackage{pgfplots}
\usetikzlibrary{positioning}
\usetikzlibrary{decorations.markings}
\usetikzlibrary{arrows,decorations.text}
\usetikzlibrary{backgrounds,fit,decorations.pathreplacing}

\makeatother

\begin{document}

\title{Enhanced realignment criterion vs. linear entanglement witnesses}


\author{Gniewomir Sarbicki}

\affiliation{Institute of Physics, Faculty of Physics, Astronomy and Informatics,
Nicolaus Copernicus University, Grudziadzka 5/7, 87-100 Toru\'{n},
Poland}

\author{Giovanni Scala}

\affiliation{Dipartimento Interateneo di Fisica, Universit\`a degli Studi di Bari,
I-70126 Bari, Italy}

\affiliation{INFN, Sezione di Bari, I-70125 Bari, Italy}

\author{Dariusz Chru\'{s}ci\'{n}ski}

\affiliation{Institute of Physics, Faculty of Physics, Astronomy and Informatics,
Nicolaus Copernicus University, Grudziadzka 5/7, 87-100 Toru\'{n},
Poland}

\date{\today}
\begin{abstract} It is shown that the enhanced (nonlinear) realignment criterion is equivalent to the  family of linear criteria based on correlation tensor. These criteria generalize the original (linear) realignment criterium and give rise to the family of entanglement witnesses.
An appropriate limiting procedure is proposed which leads to a novel class of witnesses which are as powerful as the enhanced realignment criterion.

\end{abstract}



\maketitle

\section{Introduction}

Quantum entanglement is one of key features of quantum theory. Being a fundamental theoretical concept it also provides an important resource for modern quantum technologies like quantum communication, quantum cryptography, and quantum calculations \cite{HHHH,QIT}. Let us recall that a pure state represented by a vector $\psi \in \mathcal{H}_A \otimes \mathcal{H}_B$ is separable if it has a product structure, that is, $\psi = \psi_A \otimes \psi_B$, with $\psi_A \in \mathcal{H}_A$ and $\psi_B \in \mathcal{H}_B$. For mixed states represented by density operators the definition of separable states was provided in \cite{Werner}: a bipartite state $\rho$ is separable if it allows the following decomposition
 $ \rho = \sum_k p_k \rho^A_k \otimes \rho^B_k$,
where $p_k$ is a probability distribution and $\rho^A_k$ and $\rho^B_k$ are density operators of subsystem $A$ and $B$, respectively.
There are several tools which enable one to decide whether a given state is separable or entangled \cite{GT,HHHH}.  For low dimensional bipartite systems $2\otimes 2$ (qubit-qubit) and $2 \otimes 3$ (qubit-qutrit) the celebrated Peres-Horodecki criterium states that $\rho$ is separable  if and only if it is positive partial transpose (PPT) \cite{PPT1,PPT2}.
Any entangled state $\rho_{\rm ent}$  can be detected by a suitable entanglement witness (EW),  that is, a Hermitian operator $W$ acting in $\mathcal{H}_A \otimes \mathcal{H}_B$ such that for all separable states ${\rm Tr}(W \rho_{\rm sep}) \geq 0$ but ${\rm Tr}(W \rho_{\rm ent}) < 0$ \cite{EW,HHHH,GT}(see also \cite{TOPICAL} for a recent review). This criterion is universal, that is, for any entangled state $\rho$ there exists an entanglement witness (not unique) $W$ such that ${\rm Tr}(W\rho)<0$. There is a number of other criteria \cite{GT,HHHH} which are not universal, i.e. do not allow to detect all entangled states, however they are easily applicable in practice.  The prominent example is realignment or computable cross-norm (CCNR) criterion \cite{R1,R2,R3}: if $\rho$ is separable, then
\begin{equation}\label{R}
  \| \mathcal{R}(\rho) \|_1 \leq 1 ,
\end{equation}
where $\| X \|_1 = {\rm Tr}\sqrt{X X^\dagger}$ stands for the trace norm, and $\mathcal{R}$ is a realignment operation defined as follows: if

$$ \rho = \sum_{i,j=1}^{d_A} \sum_{a,b=1}^{d_B} \rho_{ia;jb} |i\rangle \langle j| \otimes |a\rangle \langle b| , $$
then  $[\mathcal{R}(\rho)]_{ij;ab} := \rho_{ia;jb}$. Equivalently, introducing a vectorization of an operator $A = \sum_{i,j} A_{ij} |i\rangle \langle j|$ via $|A\rangle\!\rangle = \sum_{i,j} A_{ij} |i\rangle \otimes |j \rangle$ one has $\mathcal{R}(A \otimes B) = |A\rangle\!\rangle \langle\!\langle B^*|$, where the complex conjugation is taken w.r.t. the basis used for the {vectorization}. Interestingly, CCNR criterion was further generalized in \cite{ZZZ} as follows: if $\rho$ is separable, then
\begin{equation}\label{RR}
  \| \mathcal{R}(\rho -\rho_A\otimes \rho_B) \|_1 \leq \sqrt{1 - {\rm Tr}\rho_A^2} \sqrt{1 - {\rm Tr}\rho_B^2} ,
\end{equation}
where $\rho_A = {\rm Tr}_B\rho$ and $\rho_B = {\rm Tr}_A \rho$ are local states in $A$ and $B$ subsystems, respectively. Enhanced realignment criterion (\ref{RR}) turns out to be a special case of the covariant matrix criterion (CMC) \cite{COV-1,COV-2,COV-3} which was further analyzed in \cite{Lupo1,Lupo2}.

Interestingly, the enhanced criterion (\ref{RR}) is equivalent to the following {family of} nonlinear {(quadratic)} witnesses
\cite{ZZZ_EW1}

\begin{equation} \label{NEW_Q}
  W(\rho) = \mathrm{Tr}\left( \left( \oper_A \otimes \oper_B  - \sum_{\mu=0}^{d^2-1} G^A_\mu \otimes G^B_\mu \right) \rho \right) - \frac 12 \left( \mathrm{Tr}\left( \left( \sum_{\alpha=0}^{d_A^2-1} G^A_\alpha \otimes \oper_B + \sum_{\beta=0}^{d_B^2-1} \oper_A \otimes G^B_\beta \right) \rho \right) \right)^2,
\end{equation}
with $G^A_\alpha$ and $G^B_\beta$ {being} local orthonormal basis for $A$ and $B$ systems, respectively, and $d=\min\{d_A,d_B\}$. The expectation value minimal among this family for a state $\rho$ reads \cite{ZZZ_EW2}
\begin{equation}\label{NEW}
  \mathcal{F}(\rho) = 1 - \| T \|_1 - \frac 12 ({\rm Tr}\rho_A^2 + {\rm Tr}\rho_B^2) ,
\end{equation}
where $\|T\|_1$ stands for the trace norm of $d_A^2 \times d_B^2$ matrix
\begin{equation}\label{tau}
  T_{\alpha\beta} = {\rm Tr}( [\rho-\rho_A \otimes \rho_B] G^A_\alpha \otimes {G}^B_\beta).
\end{equation}
Note, that CCNR criterion (\ref{R}) may be equivalently reformulated as follows
\begin{equation}\label{}
  \| C \|_1 \leq 1 ,
\end{equation}
where the correlation tensor $C$ reads

\begin{equation}\label{}
 C_{\alpha\beta} =  {\rm Tr}(\rho G^A_\alpha \otimes {G}^B_\beta) .
\end{equation}
There are also other separability criteria based on correlation tensor \cite{Gd-1,Gd-2} which work both for bipartite and  multipartite scenario. Recently, an interesting analysis of non-linear entanglement identifiers was  performed in \cite{Laskowski}.

In this paper we show that enhanced CCNR (\ref{RR}) is equivalent to the whole family of separability criteria based on correlation tensor derived recently in \cite{RECENT}. Since separability criteria derived in \cite{RECENT} give rise to a family of (linear) entanglement witnesses, we prove that detection power of this family of witnesses is exactly the same as detection power of non-linear witnesses (\ref{NEW}).

\section{A family of separability criteria based on correlation tensor}

Let $\LH$ denotes a vector space of linear operators acting on the finite dimensional Hilbert space $\mathcal{H}$. It is endowed with the Hilbert-Schmidt inner product $\< A|B\> = {\rm Tr}(A^\dagger B)$. Now let us fix two orthonormal basis $G^A_\alpha$ and $G^B_\beta$ in $\mathrm{L}(\mathcal{H}_A)$ and $\mathrm{L}(\mathcal{H}_B)$, respectively:

$$  \< G^A_{\alpha}|G^A_{\beta}\> = \delta_{\alpha \beta}, \qquad\< G^B_\mu|G^B_\nu \> = \delta_{\mu \nu}  . $$
Note, that the trace norm $\|T \|_1 := {\rm Tr}|T|$  does not depend upon the particular orthonormal basis $G^A_\alpha$ and $G^B_\beta$. From now on we use {a} special basis such that $G^A_0 = \oper_A/\sqrt{d_A}$ and $G^B_0 = \oper_B/\sqrt{d_B}$, where $d_A = {\rm dim}\,\mathcal{H}_A$ and $d_B = {\rm dim}\, \mathcal{H}_B$. {Moreover, the remaining $G^A_\alpha$ and $G^B_\beta$ are Hermitian (and of course traceless).}

In a recent paper \cite{RECENT} we derived a family of separability criteria based on a correlation tensor $C_{\alpha\beta} = {\rm Tr}(\rho G^A_\alpha \otimes G^B_\beta)$. It was shown \cite{RECENT} that for any separable $\rho$ in $\mathcal{H}_A\otimes \mathcal{H}_B$ one has
\begin{equation}\label{xy}
  \| D^A_x C D^B_y \|_{1} \leq \mathcal{N}_{A}(x) \mathcal{N}_B(y) ,
\end{equation}
where
\begin{equation}\label{NN}
   \mathcal{N}_A(x) = \sqrt{ \frac{d_A -1 + x^2}{d_A}},\qquad \mathcal{N}_B(y)= \sqrt{ \frac{d_B -1 + y^2}{d_B}} ,
\end{equation}
for arbitrary $x,y \geq 0$, and $D^A_x = \mathrm{diag}\{x, 1, \dots, 1\}$ and $D^B_y = \mathrm{diag}\{y, 1, \dots, 1\}$ are diagonal $d_A \times d_A$ and $d_B \times d_B$ matrices, respectively. This criterion recovers several well known criteria: $(x,y)=(1,1)$ recovers original CCNR criterium,  $(x,y)=(0,0)$ recovers de Vicente result \cite{Vicente}, and $(x,y)=(\sqrt{d_A+1},\sqrt{d_B+1})$ the recent criterion based on SIC POMVs (ESIC) \cite{GUHNE}. For any fixed $(x,y)$ separability criterion (\ref{xy}) gives rise to a family of entanglement witnesses
\begin{align}  \label{W-xy}
  {W}    =   \mathcal{N}_A(x)\,  \mathcal{N}_B(y)  \, \oper_{A}\otimes \oper_{B}
  +  \sum_{\alpha=0}^{d_A^2}\sum_{\beta=0}^{d_B^2} (D^A_{x})_{\alpha \alpha}  O^{\alpha\beta} (D^B_{y})_{\beta \beta} G_\alpha^{A}  \otimes G_\beta^B
\end{align}
where $O^{\alpha\beta}$ is a real $d_A^2 \times d_B^2$ isometry.

{
\begin{Remark} Actually, one may enlarge the family replacing the set of all isometries $O^{\alpha\beta}$ by its convex hull -- a set of all real matrices $M^{\alpha\beta}$ such that $\| M\| \leq 1$, i.e. the maximal singular value is upper bounded by 1. Nevertheless such change does not increase the detection power of the family, because (from linearity of the expected value) for a given $\rho$ the minimal expectation value will be always achieved via a witness defined by an extremal element of the family -- an isometry matrix.
\end{Remark}
}

In this paper we show that enhanced CCNR (\ref{RR}) is equivalent to the whole family of criteria (\ref{xy}).



\section{Equivalence of criteria}

To prove the equivalence we start with the following

\begin{Proposition} \label{TH-1} A state $\rho$ satisfying the enhanced CCNR criterion (\ref{RR}) satisfies (\ref{xy}) for all values of parameters $x,y \geq 0$.
\end{Proposition}
Proof: Let us note that the correlation matrix $C$ for a product state is of rank one:
\begin{equation} \label{Crr}
  C(\rho_A \otimes \rho_B) =
  \left[ \begin{array}{c} \frac 1{\sqrt{d_A}} \\ \hline \mathbf{r}_A \end{array} \right]
  \left[ \begin{array}{c|c} \frac 1{\sqrt{d_B}} & \mathbf{r}_B^T \end{array} \right] ,
\end{equation}
being a product of one-particle correlation matrices. In (\ref{Crr}) $\mathbf{r}_A$ and $\mathbf{r}_B$ are Bloch vectors corresponding to $\rho_A$ and $\rho_B$, respectively, that is,

$$   \rho_A = \frac{1}{d_A} \oper_A + \sum_{\alpha>0} (\mathbf{r}_A)_\alpha G^A_\alpha  , $$
and similarly for $\mathbf{r}_B$.  One has
\begin{align}
  C(\rho)  = \left[ \begin{array}{c|c}
    \frac 1{\sqrt{d_Ad_B}} & \frac 1{\sqrt{d_A}} \mathbf{r}_B^T
    \\ \hline
    \frac 1{\sqrt{d_B}} \mathbf{r}_A & {\mathbf{C}}
  \end{array} \right]
   =
  \left[ \begin{array}{c} \frac 1{\sqrt{d_A}} \\ \hline \mathbf{r}_A \end{array} \right]
  \left[ \begin{array}{c|c} \frac 1{\sqrt{d_B}} & \mathbf{r}_B^T \end{array} \right]
  + C(\rho-\rho_A\otimes\rho_B) ,
\end{align}
and hence
\begin{align} \label{decomp}
  D^A_x C(\rho) D^B_y  & =
  \left[ \begin{array}{c} \frac x{\sqrt{d_A}} \\ \hline \mathbf{r}_A \end{array} \right]
  \left[ \begin{array}{c|c} \frac y{\sqrt{d_B}} & \mathbf{r}_B^T \end{array} \right]
  + C(\rho-\rho_A\otimes\rho_B) .
\end{align}
Let us observe that
\begin{equation}\label{purities}
 \mathrm{Tr} \rho_A^2 = \frac 1{d_A} + |\mathbf{r}_A|^2 , \quad \ \mathrm{Tr} \rho_B^2 = \frac 1{d_B} + |\mathbf{r}_B|^2 .
\end{equation}
Assume now, that the enhanced realignment criterion (\ref{RR}) is satisfied for a state $\rho$. Due to triangle inequality for the trace norm and the decomposition (\ref{decomp}) one has:
\begin{align}
  \lVert D^A_x C(\rho) D^B_y \rVert_1
  & \le \sqrt{\frac{x^2}{d_A}+|\mathbf{r}_A|^2} \sqrt{\frac{y^2}{d_B}+|\mathbf{r}_B|^2}
   + \lVert C(\rho-\rho_A\otimes\rho_B) \rVert_1
   \\
  & \le \sqrt{\frac{x^2}{d_A}+|\mathbf{r}_A|^2} \sqrt{\frac{y^2}{d_B}+|\mathbf{r}_B|^2}
   + \sqrt{1 - \frac 1{d_A} - |\mathbf{r}_A|^2}\sqrt{1 - \frac 1{d_B} - |\mathbf{r}_B|^2} \nonumber .
\end{align}
Finally, using the following property
$$ \sqrt{a}\sqrt{b}+\sqrt{c}\sqrt{d} \le \sqrt{a+c}\sqrt{b+d} $$
which holds for any non-negative $a,b,c,d$, one gets
\begin{align}
  \lVert D^A_x C(\rho) D^B_y \rVert_1 \le
   \mathcal{N}_A(x) \mathcal{N}_B(y) ,
\end{align}
which ends the proof. \hfill  $\Box$

Now, to prove the converse we find the limit of the witness (\ref{W-xy}) when $x,y \to \infty$. Formula (\ref{W-xy}) may be rewritten as follows
\begin{equation} \label{}
  {W} =  a(x,y) G^A_0 \otimes G^B_0  + x G^A_0 \otimes \sum_{\beta>0} O^{0\beta} G^B_\beta
	      +  y \sum_{\alpha>0} O^{\alpha 0} G^A_{\alpha} \otimes G^B_0 + \sum_{\alpha,\beta > 0} O^{\alpha\beta} G^A_\alpha \otimes G^B_\beta  \label {W2}
\end{equation}
where
\begin{equation}\label{}
  a(x,y) =  \sqrt{d_A -1 + x^2} \sqrt{d_B -1 + y^2} + xy\, O^{00} .
\end{equation}
Introducing polar coordinates
\begin{equation}
 x = r \cos \theta, \qquad y = r \sin \theta
\end{equation}
with $\theta \in [0,\pi/2]$, and assuming that $O^{\alpha\beta}$ does not depend on $(x,y)$ the limit $r \to \infty$ exists iff $O^{00}=-1$, and $O^{\alpha 0} = O^{0 \beta}=0$ for $\alpha,\beta > 0$, that is, $O^{\alpha\beta}$ has the following structure
\begin{equation}
  O = \left[ \begin{array}{c|c} - 1 &  \mathbf{0}^T \\ \hline  \mathbf{0} & \mathbf{O} \end{array} \right] ,
\end{equation}
where $\mathbf{O}$ is a $(d_A^2-1)\times (d_B^2-1)$ real isometry matrix.
It gives rise to the following limiting formula
\begin{align}
  {W^\infty}
  = &  a(\theta) G^A_0 \otimes G^B_0  +  \sum_{\alpha,\beta > 0} O^{\alpha\beta} G^A_\alpha \otimes G^B_\beta \label {W3}
\end{align}
with
\begin{equation}\label{}
  a(\theta) = \frac 12 \left( (d_B-1)\cot\theta + (d_A -1)\tan\theta \right) .
\end{equation}
Finally, minimizing $a(\theta)$ w.r.t. $\theta$ one finds
\begin{equation}\label{}
  a_{\rm min} = \sqrt{(d_A-1)(d_B-1)} ,
\end{equation}
which reproduces EW corresponding to de Vicente criterion \cite{Vicente}. To get more refined limit let us assume that $O^{\alpha\beta}$ can depend on $(x,y)$.  The only way to guarantee the existence of the limit $r\to \infty$ is to assume the following asymptotics for the matrix elements of an isometry $O^{\alpha\beta}$
\begin{equation}
  O^{00} = - \sqrt{1 - \frac {\eta^2}{r^2}} + O(r^{-2}) \label{limit_ass}
\end{equation}
together with
\begin{equation}\label{}
  O^{0\beta} = \frac \eta r v^\beta + O(r^{-2}), \qquad  O^{\alpha0} = \frac \eta r  u^\alpha +O(r^{-2}) ,
\end{equation}
for $\alpha, \beta > 0$, where $\mathbf{u} \in \mathbb{R}^{d_A^2-1}$ and $\mathbf{v} \in \mathbb{R}^{d_B^2-1}$. One finds in the limit $r \to \infty$
\begin{equation} \label{W4}
 {W^\infty}
  =  b(\theta,\eta) G^A_0 \otimes G^B_0 +  \sum_{\alpha,\beta > 0} O^{\alpha\beta} G^A_\alpha \otimes G^B_\beta
  + \eta \Big( \cos\theta G^A_0 \otimes \sum_{\beta>0} v^{\beta} G^B_\beta
  + \sin\theta \sum_{\alpha>0} u^{\alpha} G^A_{\alpha} \otimes G^B_0 \Big)   ,
\end{equation}
with
\begin{equation*}\label{}
  b(\theta, \eta) =  \frac 12 \Big( (d_B-1)\cot\theta + (d_A -1)\tan\theta + \eta^2 \sin\theta\cos\theta \Big) .
\end{equation*}

The isometry $O^{\alpha\beta}$ has the following asymptotic structure
(up to leading powers of $1/r$)
\begin{equation}
  O^{\alpha\beta}(r) = \left[ \begin{array}{c|c} - \sqrt{1 - \frac {\eta^2}{r^2}} & \frac{\eta}{r} \mathbf{v}^T \\ \hline \frac{\eta}{r} \mathbf{u} & \sqrt{1 - \frac{\eta^2}{r^2}} \mathbf{O} \end{array} \right] ,
\end{equation}
where $\mathbf{O}$ is a $(d_A^2-1)\times (d_B^2-1)$ real matrix. Now, the isometry condition for $O^{\alpha\beta}$ imply that $\mathbf{O} \mathbf{O}^T$ and  $\mathbf{O}^T \mathbf{O}$ are ${\rm min}\{ d^2_A, d^2_B\}$--dimensional projectors and hence
 $ | \mathbf{u}| = |{\mathbf{v}}| = 1$,
together with the following constraint for $\mathbf{u}$ and $\mathbf{v}$:
\begin{equation}\label{uOv}
  \mathbf{u} = \mathbf{O} \mathbf{v} . 
\end{equation}

Summarising, the asymptotic  witness $W^\infty$ is characterized by an isometry $\mathbf{O}$, two normalized vectors satisfying (\ref{uOv}), an angle $\theta \in [0,\pi/2]$, and an arbitrary real parameter $\eta \geq 0$. Actually, one can assume that $\eta \geq 0$ since $\eta$ always multiplies $\mathbf{u}$ and $\mathbf{v}$.

Note, that in the limit $\eta \to 0$ one recovers again a witness corresponding to de Vicente criterium \cite{Vicente}.

\begin{Remark} Note, that if one  replaces an isometry $O^{\alpha\beta}$ by an arbitrary real matrix $M^{\alpha\beta}$ such that $\| M\| \leq 1$,
then one can essentially repeat all the steps of the proof and finds



\begin{equation} \label{}
  {W^\infty}
  =  b(\theta,\eta) G^A_0 \otimes G^B_0 +  \sum_{\alpha,\beta > 0} \mathbf{M}^{\alpha\beta} G^A_\alpha \otimes G^B_\beta
  + \cos\theta G^A_0 \otimes \sum_{\beta>0} v^{\beta} G^B_\beta
  + \sin\theta \sum_{\alpha>0} u^{\alpha} G^A_{\alpha} \otimes G^B_0    ,
\end{equation}
where $ \mathbf{u} = \mathbf{M} \mathbf{v}$, and $\mathbf{M}^{\alpha\beta} := M^{\alpha\beta}$ for $\alpha,\beta>0$.  Note, that $\mathbf{u}$ and $\mathbf{v}$ are no longer normalized.
\end{Remark}



{
\begin{Proposition}
  An {entangled} state detected by the enhanced CCNR criterion (\ref{RR}) is also detected by the criterion (\ref{xy}) for some values of parameters $(x,y)$.
\end{Proposition}
}

Proof:  Let us consider  an arbitrary  state $\rho$ in $\mathbb{C}^{d_A} \otimes \mathbb{C}^{d_B}$
\begin{eqnarray}
  \rho = \frac{1}{d_A d_B} \oper_A \otimes \oper_B + \widetilde{\rho} \ ,
\end{eqnarray}
where the traceless part $\widetilde{\rho}$ reads
\begin{equation}\label{}
 \widetilde{\rho} =  \frac 1{d_A} \oper_A \otimes \widetilde{\rho}_B + \widetilde{\rho}_A \otimes \frac 1{d_B} \oper_B
	    + \sum_{\alpha,\beta>0} C_{\alpha\beta} G^A_\alpha \otimes G^B_\beta
\end{equation}
with
\begin{equation*}\label{}
  \widetilde{\rho}_A = \sum_{\alpha>0} (\mathbf{r}_A)_{\alpha} G^A_\alpha , \qquad
  \widetilde{\rho}_B = \sum_{\beta>0} (\mathbf{r}_B)_{\beta} G^B_\beta .
\end{equation*}
One finds
\begin{eqnarray} \label{TR}
  {\rm Tr}
  ({W^\infty}\rho)
  &=& \frac{b(\theta,\eta)}{{\sqrt{d_A d_B}}} +  \langle \mathbf{O} | \mathbf{C} \rangle  +  \eta \left(  \frac{\cos\theta}{\sqrt{d_A}} \langle \mathbf{r}_B |  \mathbf{v}\rangle
  + \frac{\sin\theta}{\sqrt{d_B}} \langle  \mathbf{r}_A | \mathbf{u} \rangle \right) \nonumber \\
   &=& {\frac{b(\theta,\eta)}{\sqrt{d_A d_B}} +  \langle \mathbf{O} | \mathbf{C} \rangle}  {+} {\eta \langle \frac{\cos\theta}{\sqrt{d_A}} \mathbf{r}_B
    + \frac{\sin\theta}{\sqrt{d_B}} \mathbf{O}^T\mathbf{r}_A | \mathbf{v} \rangle},
\end{eqnarray}
where $\mathbf{C}_{\alpha\beta} = C_{\alpha\beta}$ for $\alpha,\beta >0$.

\begin{Lemma} For a given bipartite state $\rho$ there exists $\mathbf{u}$, $\mathbf{v}$, $\eta$, and isometry  $\mathbf{O}$ such that the corresponding witness $W^\infty$ satisfies

\begin{eqnarray}\label{TR-3a}
  {\rm Tr}(
{W^\infty}
  \rho)  = \sqrt{ (1-{\rm Tr}\rho_A^2)(1-{\rm Tr}\rho_B^2)} - \|\mathcal{R}(\rho - \rho_A \otimes \rho_B)\|_1 {,}
\end{eqnarray}
and this is the minimal value of $ {\rm Tr}({W^\infty}\rho)$ for a given state $\rho$.
\end{Lemma}
Proof of the lemma: observe that to minimise
$ {\rm Tr}({W^\infty}\rho) $
the unit vector $\mathbf{v}$ has to be antiparallel to $\frac{\eta\cos\theta}{\sqrt{d_A}} \mathbf{r}_B + \frac{\eta\sin\theta}{\sqrt{d_B}} \mathbf{O}^T \mathbf{r}_A$, where we used $\mathbf{u} = \mathbf{O}\mathbf{v}$.  The {third} summand in (\ref{TR})  becomes then $- \eta \left| \frac{\cos\theta}{\sqrt{d_A}} \mathbf{r}_B + \frac{\sin\theta}{\sqrt{d_B}} \mathbf{O}^T \mathbf{r}_A \right|$. Let us perform now minimisation w.r.t.  parameter $\eta$. One easily finds
\begin{equation}\label{eta-min}
  {\eta_{\rm min}} =  \left| \frac{\cos\theta}{\sqrt{d_A}} \mathbf{r}_B + \frac{\sin\theta}{\sqrt{d_B}} \mathbf{O}^T \mathbf{r}_A \right| \frac{\sqrt{d_Ad_B}}{{\sin\theta\cos\theta}} ,
\end{equation}
and hence for these particular parameters the value of ${\rm Tr}(W^\infty \rho)$ reads
{
\begin{eqnarray} \label{TRa}
  {\rm Tr}(
  {W^\infty}
  \rho)
  & = & \frac{(d_B-1)\cot\theta + (d_A -1)\tan\theta}{2 \sqrt{d_A d_B}}  + \langle \mathbf{O} | \mathbf{C}\rangle
  -  \left| \frac{\cos\theta}{\sqrt{d_A}} \mathbf{r}_B + \frac{\sin\theta}{\sqrt{d_B}} \mathbf{O}^T \mathbf{r}_A \right|^2 \frac{\sqrt{d_Ad_B}}{ 2 \sin\theta\cos\theta}
  \nonumber \\
  &=& \frac{(d_B-1)\cot\theta + (d_A -1)\tan\theta}{2 \sqrt{d_A d_B}}
  - \sqrt{d_Ad_B}\left( \frac{\cot\theta}{2d_A} |\mathbf{r}_B|^2 + \frac{\tan\theta}{2d_B} |\mathbf{r}_A|^2 \right)
 +  \langle \mathbf{O} | \mathbf{C} \rangle + \langle \mathbf{r}_B | \mathbf{O}^T \mathbf{r}_A \rangle
  \nonumber \\
  &=& \frac 1{2\sqrt{d_Ad_B}} ( \cot\theta \left( d_B - 1 - d_B |\mathbf{r}_B|^2\right)
  + \tan\theta \left( d_A - 1 - d_A |\mathbf{r}_A|^2\right) )
  + \langle \mathbf{O} | \mathbf{C} - \mathbf{r}_A \mathbf{r}_B^T \rangle \nonumber {.}
\end{eqnarray}
}
Finally, using the following identities from Eq. (\ref{purities})
$$   1 - {\rm Tr}\rho_A^2 = \frac {1}{d_A}\left( d_A-1 - d_A|\mathbf{r}_A|^2\right), \qquad   1 - {\rm Tr}\rho_B^2 = \frac {1}{d_B}\left( d_B-1 - d_B|\mathbf{r}_B|^2\right) , $$
one finds
\begin{eqnarray} \label{TR-2}
  {\rm Tr}(
  {W^\infty}
  \rho)   =  \frac{d_B(1-{\rm Tr}\rho_B^2)\cot\theta + d_A(1- {\rm Tr}\rho_A^2)\tan\theta}{2 \sqrt{d_A d_B}}  +  \langle \mathbf{O} | \mathbf{T} \rangle ,
\end{eqnarray}
where $\mathbf{T}_{\alpha\beta} = T_{\alpha\beta}$ (from Eq. (\ref{tau})) for $\alpha,\beta >0$, that is,
$$   \mathbf{T}_{\alpha\beta} = \mathbf{C}_{\alpha\beta} - (\mathbf{r}_A)_\alpha (\mathbf{r}_B)_\beta . $$
The last step is {the} minimization w.r.t. $\theta$ and {the} isometry $\mathbf{O}$. One finds for the optimal $\theta$
\begin{equation}\label{th-min}
  \tan\theta_{\rm min} = \sqrt{\frac{d_B(1- {\rm Tr}\rho_B^2)}{d_A(1- {\rm Tr}\rho_A^2)}}
\end{equation}
and
\begin{equation}\label{}
  \min_\mathbf{O} \< \mathbf{O}|\mathbf{T}\> = - \max_{\mathbf{O}}  \< \mathbf{O}|\mathbf{T}\>  =- \lVert \mathbf{T} \rVert_1,
\end{equation}
and hence noting that $\< \mathbf{O}|\mathbf{T}\> = \< {O}|{T}\>$ one finally arrives at (\ref{TR-3a}). \hfill $\Box$

\vspace{.1cm}

Clearly, if
$\rho$ is detected by  the enhanced CCNR criterion, then due to the Lemma one can find a witness $W^\infty$ detecting $\rho$ as well.  While the witness   {$W^\infty$}   is realised as a limit of witnesses ${W}$ (\ref{W-xy}), there exist witnesses ${W}$ detecting the state for  large enough $x$ and $y$, which ends the proof. \hfill $\Box$

\vspace{.1cm}

In summary we proved the following

\begin{Theorem} Enhanced realignment criterion (\ref{RR}) is equivalent to the family of (linear) entanglement witnesses (\ref{W-xy}).
\end{Theorem}

Interestingly, our analysis enables one to construct a witness for an entangled state detected by (\ref{RR}). Indeed, observe that $T_{00}=T_{0\beta} = T_{\alpha0}=0$ and hence the entire information of $T$ is encoded into $\mathbf{T}$. Now, consider a singular value decomposition
$$    \mathbf{T} = \mathbf{O}_1 \mathbf{D} \mathbf{O}_2^T , $$
with $\mathbf{O}_1$ and $\mathbf{O}_2$ orthogonal matrices and let {$\mathbf{O} := \mathbf{O}_1 \mathbf{O}_2^{T}$}. The corresponding angle $\theta$ is defined in (\ref{th-min}) and the parameter $\eta$ is defined in (\ref{eta-min}). Finally, a unit vector $\mathbf{v}$ reads
\begin{equation}\label{}
  \mathbf{v} = - \frac{ \frac{\cos\theta}{\sqrt{d_A}} \mathbf{r}_B + \frac{\sin\theta}{\sqrt{d_B}} \mathbf{O}^T \mathbf{r}_A \ }{\left| \frac{\cos\theta}{\sqrt{d_A}} \mathbf{r}_B + \frac{\sin\theta}{\sqrt{d_B}} \mathbf{O}^T \mathbf{r}_A \right| } ,
\end{equation}
and it is fully determined by $\mathbf{r}_A$, $\mathbf{r}_B$, the isometry $\mathbf{O}$, and the angle $\theta$.

\section{Conclusions}

In conclusion we shown that the enhanced realignment criterion (\ref{RR}) which is nonlinear in $\rho$ is perfectly equivalent to a family of linear criteria based on (\ref{xy}).
These criteria are equivalent to a family of entanglement witnesses derived recently in \cite{RECENT}. Here we derived a limit $x,y\to \infty$ which gives rise to a novel class of entanglement witnesses. Interestingly, it is shown that given an entangled state detected by the enhanced realignment criterion one is able to construct a witness from our new class which detects entanglement of this state as well.

The enhanced realignment criterion is a powerful tool for detecting entanglement in bipartite systems. Clearly, this criterion is not universal and there are quantum entangled states which are not detected by this criterion. An interesting class of states was recently considered in \cite{DS}: so-called Diagonal Symmetric (DS) states in $\mathbb{C}^d \otimes \mathbb{C}^d$ defined as follows

\begin{equation}\label{DS}
  \rho = \sum_{i,j=1}^d p_{ij} |D_{ij}\rangle \langle D_{ij}| ,
\end{equation}
where $ |D_{ii}\rangle =  |i \otimes i\rangle$, $ |D_{ij}\rangle = ( |{i \otimes j}\rangle +  |{j\otimes i}\rangle)/\sqrt{2}$ for $i \neq j$, and $p_{ij}$ is a probability distribution. One checks by direct calculation that enhanced realignment criterion fails to detect entanglement of (\ref{DS}).

Our results call also for a multipartite generalization which we postpone for a future research. Actually, multipartite case was already initiated in \cite{ZZZ}. However, authors of \cite{ZZZ} considering a general multipartite case studied only entanglement of various bi-partitions of the multipartite scenario.

It would be also interesting to further analyse the current class of witnesses derived in this paper. In particular one may ask which of them are optimal and not decomposable.

Moreover, it is known in \cite{COV-1} that appropriate \emph{local filtering operations} might improve separability criteria paving the way towards future developments for the  entanglement detection method presented in this paper.

\begin{acknowledgments}
DC and GSa were supported by the Polish National Science
Centre project 2015/19/B/ST1/03095. GSc is supported by Instituto Nazionale di Fisica Nucleare (INFN) through the project ``QUANTUM''.  We acknowledge the Toru\'n Astrophysics/Physics Summer Program TAPS 2018 and the project PROM at the Nicolaus Copernicus University.
\end{acknowledgments}

\bibliographystyle{unsrtnat}

\end{document}